# Ultrahigh quality infrared polaritonic resonators based on bottom-up-synthesized van der Waals nanoribbons

Shang-Jie Yu[1,7], Yue Jiang[2,7], John A. Roberts[3], Markus A. Huber[3], Helen Yao[4], Xinjian Shi[2], Hans A. Bechtel[5], Stephanie N. Gilbert Corder[5], Tony F. Heinz[3,6], Xiaolin Zheng[2], and Jonathan A. Fan[1]*

[1]Department of Electrical Engineering, Stanford University, Stanford, California 94305, United States.

[2]Department of Mechanical Engineering, Stanford University, Stanford, California 94305, United States.

[3]Department of Applied Physics, Stanford University, Stanford, California 94305, United States.

[4]Department of Material Science and Engineering, Stanford University, Stanford, California 94305, United States.

[5]Advanced Light Source Division, Lawrence Berkeley National Laboratory, Berkeley, California 94720, United States.

[6]SLAC National Accelerator Laboratory, Menlo Park, California 94305, United States.

[7]These authors contributed equally to this work.

*To whom correspondence should be addressed (jonfan@stanford.edu)

## ABSTRACT

van der Waals nanomaterials supporting phonon polariton quasiparticles possess unprecedented light confinement capabilities, making them ideal systems for molecular sensing, thermal emission, and subwavelength imaging applications, but they require defect-free crystallinity and nanostructured form factors to fully showcase these capabilities. We introduce bottom-up-synthesized α-$MoO_3$ structures as nanoscale phonon polaritonic systems that feature tailorable morphologies and crystal qualities consistent with bulk single crystals. α-$MoO_3$ nanoribbons serve as low-loss hyperbolic Fabry-Pérot nanoresonators, and we experimentally map hyperbolic resonances over four Reststrahlen bands spanning the far- and mid-infrared spectral range, including resonance modes beyond the tenth order. The measured quality factors are the highest from phonon polaritonic van der Waals structures to date. We anticipate that bottom-up-synthesized polaritonic van der Waals nanostructures will serve as an enabling high-performance and low-loss platform for infrared optical and optoelectronic applications.

KEYWORDS: nanophotonics, polaritonics, nanoresonators, van der Waals materials, near-field optical microscopy, infrared nanospectroscopy

High-performance, low-loss photonic materials are enabling platforms for the scientific study of optical phenomena and the technological advancement of waveguide- and resonator-based systems. In the mid-infrared (MIR) and terahertz wavelength range, van der Waals (vdW) materials have emerged as an exciting and heavily studied class of materials due to their ability to support exceptionally high-quality phonon polariton (PhP) excitations, which occur in the Reststrahlen bands (RBs) between the longitudinal optical and corresponding transverse optical phonon frequencies.[1] These vdW polaritonic systems include h-BN (ref. [2]), α-$MoO_3$ (ref. [3–5]) and α-$V_2O_5$ (ref. [6]), and they have been a testbed for evaluating light-matter interactions ranging from hyperbolic dispersion to strong coupling phenomena.[2,7] They are also capable of compressing wavelengths to length scales tens to hundreds of times smaller than those in free space,[8,3] which has inspired new concepts and regimes in sensing, imaging and waveguiding.[9–11] α-$MoO_3$ is a particularly unique vdW phonon polaritonic material as it exhibits biaxial anisotropy, which enables IR polarization conversion[12] and in-plane hyperbolic PhP modes.[3] More recently, topological photonic dispersion engineering was demonstrated in twisted bilayer α-$MoO_3$ structures,[13,14] providing a new avenue towards flexible polaritonic tuning and routing.

The gold standard approach for studying and implementing low-loss vdW PhP devices is to mechanically exfoliate bulk single crystals and characterize individual flakes. While this approach is sufficient for certain proof-of-concept experiments, it is time-consuming, low-yield, and not scalable. In addition, flake-based structures that localize light using geometric structuring, such as antennas, resonators, and waveguides, require top-down patterning and etching steps that introduce damage and contamination,[15–21] precluding the study of high-performance, low-loss devices.

In this Article, we report the growth, assembly, and characterization of ultrahigh quality polaritonic systems based on α-MoO$_3$ microplates and nanoribbons, synthesized using the flame vapor deposition (FVD) method.[22,23] FVD is a versatile platform for growing nanostructured metal oxides and is noteworthy for its rapid growth rates, low cost, high scalability, and use of atmospheric operation conditions.[23,24] In prior work, flame synthesis was used to produce metal oxide nanostructures for catalysis, energy storage, and electronic doping applications.[25–28] This study represents the first IR optical analysis of α-MoO$_3$ synthesized by FVD,[26,29] and it combines multiple features in materials preparation that enable unprecedented optical properties. First, FVD-grown structures are crystalline with minimal impurities,[23,26] and they have material properties consistent with flakes exfoliated from bulk single crystals. Second, α-MoO$_3$ nanoribbons have smooth and parallel edges, naturally forming ideal Fabry-Pérot (FP) phonon polaritonic cavities without the need for nanopatterning. Third, as-grown structures can be easily transferred to any solid substrate with minimal contamination. We transfer α-MoO$_3$ samples to ultrasmooth gold substrates and for the first time cleanly map the PhP resonances across the far- and mid-infrared range with ultrabroadband IR nanospectroscopy. The PhP resonators exhibit quality factors (Q-factors) that are the highest reported in PhP resonators to date.[17,30–32] We anticipate that FVD will serve as a general and versatile growth platform for the preparation, characterization, and prototyping of high-quality optical resonators based on metal oxide nanostructures.

RESULTS AND DISCUSSION

**Flame synthesis and near-field imaging of α-MoO₃ structures**

We synthesize and prepare α-MoO₃ nano- and micro-structures using the workflow illustrated in Figure 1a. Using FVD, we grow stoichiometric and single-crystalline α-MoO₃ structures on a silicon substrate placed above a heated molybdenum source on top of a premixed methane-air flame.[29] The typical growth time is only five minutes. The as-grown α-MoO₃ is then transferred to a flat target substrate using low-adhesion tape and probed by scattering-type scanning near-field optical microscopy (s-SNOM), which is an established tool for infrared PhP study.[33] The growth and transfer substrates each have centimeter-scale regions of α-MoO₃ nano- and micro-structure coverage, with area limitations set only by the burner diameter and tape size.

The synthesis of α-MoO₃ under fuel-lean flame conditions yields thermodynamically favorable conditions for crystalline, stoichiometric materials growth(see Methods).[34] The composition of the as-grown material is confirmed to be MoO₃ with negligible oxygen deficiency using X-ray photoelectron spectroscopy, and the high degree of crystallinity is confirmed with X-ray diffraction (Supporting Information). We also probe the crystallinity and phase of as-grown MoO₃ structures using high-resolution transmission electron microscopy (HRTEM) and selected area electron diffraction (SAED), which confirm that the structures are crystalline orthorhombic α-MoO₃, the preferred growth direction is [001], and the out-of-plane orientation is [010] (Figure 1b).

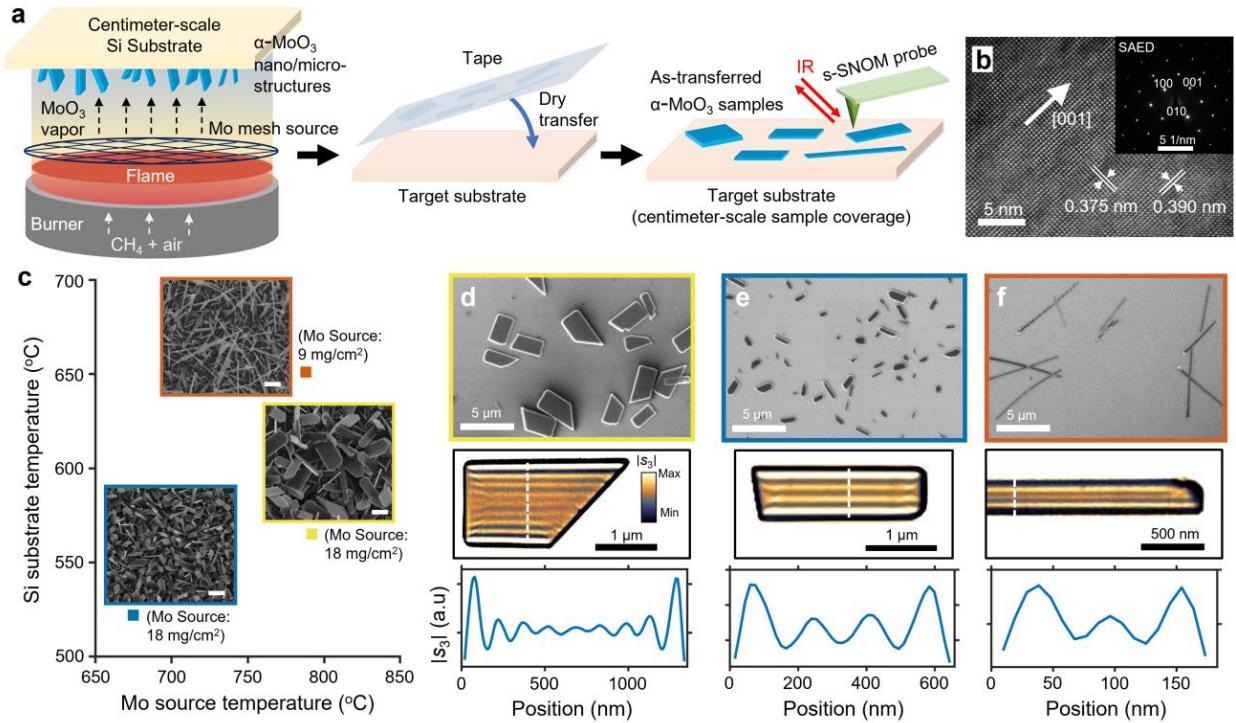

**Fig. 1. α-MoO$_3$ sample preparation, morphology control, and characterization. a**, FVD setup and materials preparation workflow, including flame synthesis, direct dry transfer, and near-field optical characterization of as-transferred α-MoO$_3$ structures. **b**, HRTEM image of a representative α-MoO$_3$ structure. Inset: SAED pattern. **c**, Morphology control of α-MoO$_3$ nano- and micro-structures by varying the FVD conditions. SEM images of as-grown α-MoO$_3$ are shown for each synthesis condition. Scale bars: 1 μm. **d-f**, Top row: SEM images of as-transferred samples on SiO$_2$/Si substrates prepared under different FVD conditions. Yellow, dark blue and dark red boxes indicate (**d**) microplates, (**e**) low-aspect-ratio nanoribbons, and (**f**) high-aspect-ratio nanoribbons, respectively. Middle row: corresponding s-SNOM images of an individual microplate and nanoribbons recorded at 931 cm$^{-1}$. Bottom row: near-field amplitude ($|s_3|$) profiles taken from the cross-sections of the middle-row images, denoted by the dashed lines.

Adjustments of the FVD growth conditions enable control of the α-MoO$_3$ structure morphology, providing a straightforward pathway to preparing different classes of materials for optical study. To control the α-MoO$_3$ morphology while preserving its stoichiometry, we fix the fuel-to-air equivalence ratio of the premixed methane flame at 0.88 and vary the substrate and molybdenum source temperature and source area density. With these temperature and area density variations, we can reproducibly synthesize α-MoO$_3$ with three different morphologies (Figure 1c): microplates (length of 2–5 μm and a width of 0.5–2.5 μm), low-aspect-ratio nanoribbons (length of 1–2.5 μm and width of 0.5–1 μm), and high-aspect-ratio nanoribbons (length of 4–10 μm and width of 0.1–0.3 μm). Their typical thickness is between 50 – 250 nm.

s-SNOM images of individual α-MoO$_3$ structures grown under different conditions (Figure 1d-f) and taken within RB$_Y$ (821.5 – 962.9 cm$^{-1}$ from phonons along [001]) show distinct regimes of PhP behavior. The microplates, which have widths longer than the PhP propagation distance, show long-range PhP propagation with exponentially decaying envelopes. The nanoribbons, on the other hand, have widths much shorter than the PhP propagation distance and serve as FP resonators that support PhP standing waves.[35,36] The resonances satisfy the FP condition $2\text{Re}(q)L + 2\varphi = 2\pi n$, where $q$ is the in-plane PhP mode wavevector, $L$ is the resonator size, $\varphi$ is the reflection phase accumulated by the PhP modes at the ribbon edge, and $n$ is an integer indicating mode order. The FP resonance order can be tailored by selecting nanoribbons with differing thicknesses and widths. In the following sections, we will use the microplates to quantify PhP propagation characteristics and the nanoribbons to characterize PhP-based resonator behavior.

**PhP mode dispersion and figure of merit in microplates**

To quantify the quality of the phonon resonances in FVD-grown α-MoO$_3$ microplates with optical metrology methods, we utilize a combination of Raman spectroscopy and s-SNOM methods. With Raman spectroscopy, we can directly probe the in-plane anisotropic Raman response and extract the phonon lifetimes from the peak width. Representative Raman spectra for structures on oxidized silicon substrates are plotted in Figure 2a and display a strong dependence on the pump light polarization with respect to the ribbon crystal axis at 160 cm$^{-1}$, which is attributed to anisotropic lattice vibrations within the vdW α-MoO$_3$ plane. This result is consistent with previous work[37] and confirms the direction of ribbon growth. We also quantify the phonon lifetimes of Raman active modes at 995 cm$^{-1}$ and 818 cm$^{-1}$ from the microplates and compare them with exfoliated flakes from bulk single crystals. The results from this benchmark comparison are summarized in Figure 2b and show that the lifetimes from both materials are similar and consistent with previously published data on exfoliated flakes.[3] The PhP lifetimes from FVD-grown and exfoliated samples are also quantified by s-SNOM and produce values consistent with the Raman analysis in Figure 2c (see Supporting Information). All these results indicate that our FVD-grown structures have comparable crystal quality with ideal single crystal samples prepared by bulk exfoliation.

To further engineer the α-MoO$_3$ PhP propagation mode in a scalable manner that eliminates phonon-polariton interactions with the underlying substrate, we tape-transfer α-MoO$_3$ microplates onto freshly template-stripped ultrasmooth gold substrates (Figure 3a). Our use of ultrasmooth gold substrates offers several advantages. First, the elimination of phonon contributions from the underlying substrate removes ambiguity involving the source of PhPs across the full infrared frequency range and mitigates modal absorption from coupling to substrate phonons. While gold does introduce free carrier absorption, this absorption contribution is minimal at mid- and far-infrared (FIR) wavelengths in our system (see Supporting Information). Second, freshly template-stripped gold has surface roughness on the angstrom scale and therefore contributes negligible surface scattering, which is critical to minimizing losses during PhP propagation. Evaporated or sputtered gold substrates, which have been previously used in h-BN device studies,[38,39] are rougher and contribute greater surface scattering loss. Third, the gold substrate at infrared frequencies enforces mirror-like electromagnetic boundary conditions at the gold-microplate interface, leading to minimal field penetration into the substrate and strong tailoring of mode confinement and mode symmetry properties.

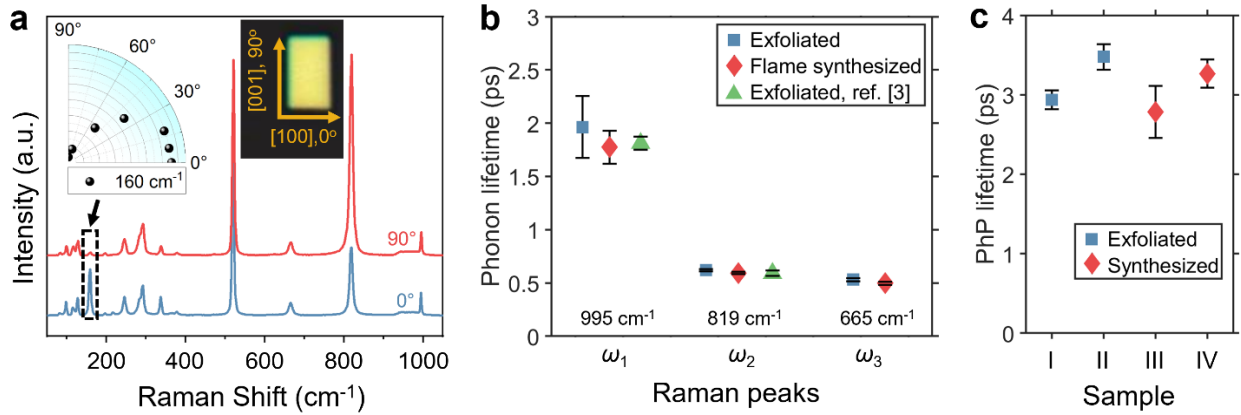

**Fig. 2. Raman spectroscopy and phonon lifetime. a**, Raman spectra of a microplate with the excitation light polarized along the short edge (0°) and the long edge (90°). Left-top inset: the excitation polarization dependence of Raman peak at 160 cm$^{-1}$ showing in-plane anisotropy. Right-top inset: optical image of the microplate characterized by Raman spectroscopy. The arrows denote the crystal axes and polarization angle. **b**, Phonon lifetimes of exfoliated and flame synthesized samples, extracted from the Raman spectra. The first dataset is collected on 12 exfoliated flakes, the second dataset is on 22 flame-synthesized microplates and nanoribbons, and the third dataset is on 6 exfoliated flakes from Ref. [3]. The labels $\omega_1$, $\omega_2$, and $\omega_3$ correspond to the phonon lifetime data of the Raman peaks at 995 cm$^{-1}$, 819 cm$^{-1}$, and 665 cm$^{-1}$, respectively. Data from the 665 cm$^{-1}$ peak are not available for the third dataset. The error bars denote standard deviations for phonon lifetime evaluation from Raman peak fitting. **c**, PhP lifetimes of exfoliated and flame synthesized samples, extracted from s-SNOM imaging at 939 cm$^{-1}$. The error bars show the standard deviations from fitting.

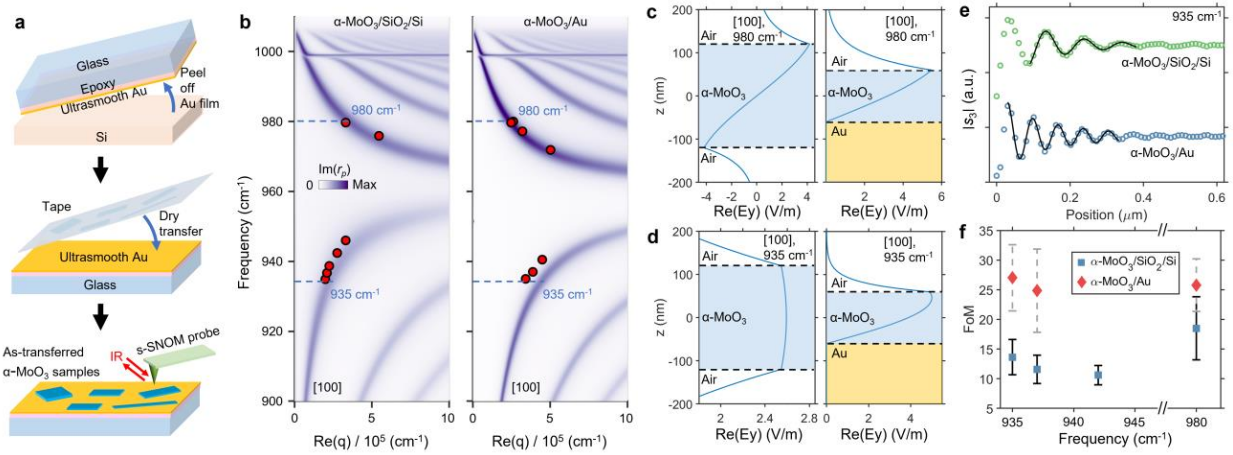

**Fig. 3. PhP mode dispersion and figure of merit**. **a**, Material preparation workflow, including ultrasmooth gold template stripping, direct dry transfer, and s-SNOM characterization of as-transferred α-MoO$_3$ samples. **b**, Left: PhP dispersion of 112 nm-thick α-MoO$_3$ on SiO$_2$, with in-plane momentum $q$ along [001] (top) and along [100] (bottom). The points are experimental data for microplates on 90 nm SiO$_2$/Si. Right: PhP dispersion of 120 nm-thick α-MoO$_3$ on gold, with $q$ along [001] (top) and along [100] (bottom). The points are the corresponding experimental data on ultrasmooth gold. **c-d**, Calculated electric field profiles in the vertical cross-sections of guided modes along [100] direction in air/240 nm thick α-MoO$_3$/air and air/120 nm thick α-MoO$_3$/gold at (**c**) 980 cm$^{-1}$ and (**d**) 935 cm$^{-1}$. The y-axis is in the out-of-plane direction (*i.e.*, normal to plot plane). **e**, Typical s-SNOM fringe traces on α-MoO$_3$/SiO$_2$/Si and α-MoO$_3$/ultrasmooth gold samples imaged at 935 cm$^{-1}$. The circles are experimental data and the black lines are fitting curves. **f**, Figure of merit (FoM), defined as Re($q$)/Im($q$), of PhPs measured from various microplates on SiO$_2$ and ultrasmooth gold, evaluated by fitting s-SNOM fringe traces.

We study the dispersion of α-MoO$_3$ microplates on ultrasmooth metal substrates at the RB$_Y$ and RB$_{Z2}$ bands and compare the results with microplates on SiO$_2$/Si substrates. The PhP momenta are extracted through s-SNOM imaging. We also calculate theoretical dispersion plots using permittivity values published elsewhere.[40,41] The results (Figure 3b) show good agreement between experimental and theoretical curves and display clear spatial mode compression near the

α-MoO$_3$ phonon resonance, with wavelength compression as large as λ/50. Notably, as the substrate changes from SiO$_2$/Si to gold, the upper RB dispersion curve shifts towards lower momentum while the lower RB dispersion curve shifts towards higher momentum. This dispersion dependence on substrate material can be understood in the context of mode symmetry. In the metal-substrate system, the gold surface enforces an electric field node at the α-MoO$_3$-metal interface, such that the electromagnetic modes take the form of anti-symmetric modes in an air-clad α-MoO$_3$ film with a doubled thickness (Figure 3c and 3d). The enforcement of anti-symmetry in the mode profiles therefore serves as a filter that eliminates the presence of symmetric modes in the air-clad α-MoO$_3$ picture. A more detailed discussion can be found in the Supporting Section.

We also experimentally quantify the figure of merit (FoM) of the intrinsic PhP mode quality, defined as Re($q$)/Im($q$), by processing a series of microplate s-SNOM scans imaged at different wavelengths. The FoM is independent of the thickness (see Supporting Information), allowing direct comparisons among microplates with different thicknesses. s-SNOM scans for representative α-MoO$_3$ microplates on the two different substrates, measured at 935 cm$^{-1}$ in the lower RB, are shown in Figure 3e. The microplates on SiO$_2$/Si support PhPs with a FoM = 11 while the samples on ultrasmooth gold support PhPs with a further reduced wavelength, due to the mode tailoring discussed above, and a FoM = 29. The enhanced FoM in the gold-based systems is due to reduced environmental loss in the substrate and enhanced mode compression. A more thorough quantification of the FoM for different samples and wavelengths is summarized in Figure 3f and shows that PhPs supported by α-MoO$_3$ microplates on gold have consistently high FoMs in both the upper and lower RBs.

**Nanospectroscopy of high-Q PhP nanoribbon resonators**

We next study the FP PhP modes within the RBs of an α-MoO$_3$ nanostructure using FIR and MIR ultrabroadband synchrotron infrared nanospectroscopy (SINS).[42–44] FIR SINS covers the four α-MoO$_3$ RBs spanning 440 – 1020 cm$^{-1}$ (Figure 4a) and MIR SINS enables scans in RB$_{Z2}$ with high spectral resolution due to its higher MIR photon detection sensitivity (see details in Methods). For this analysis, we probe a rectangular nanoribbon on a gold substrate (AFM image in Figure 4b and s-SNOM image at 980 cm$^{-1}$ in Figure 4c), which supports FP resonances across both axes and possesses no substrate phonon contributions across the full wavelength range of interest. The SINS linescans over the nanoribbon reveal clear FP resonances over multiple orders along the [001] and [100] directions (Figure 4d-e, linescans indicated in Figure 4b). In RB$_X$ (RB$_Y$), the map displays resonances only along [001] ([100]), and in RB$_{Z1}$ and RB$_{Z2}$, there are resonances along both axes. These resonances are much stronger and more discernable than those from nanostructures grown using alternative material preparation methods, such as the solution growth of α-MoO$_3$ nanostructures.[45]

We simulate the hyperspectral maps by scanning an out-of-plane polarized dipole along the crystal axis in a finite-difference time-domain simulator, and the resulting scans (Figure 4f) agree well with the experiment. Differences between the experimental and calculated spectra come from two primary sources. First, the permittivity parameters used in the simulation utilize literature values[40,41] that are taken from samples prepared differently than those in our study. Second, the simulated maps are based on a 2D model and only capture FP resonances across one lateral dimension of the resonator, whereas an experimental spectrum combines features from both x- and y-dimensions. The simulated hyperspectral scans for a full 3D system show that a dipole can simultaneously excite resonances across both lateral dimensions, and furthermore, the superposition of both contributions can blur resonant features, particularly those involving high-order FP modes (see Supporting Information).

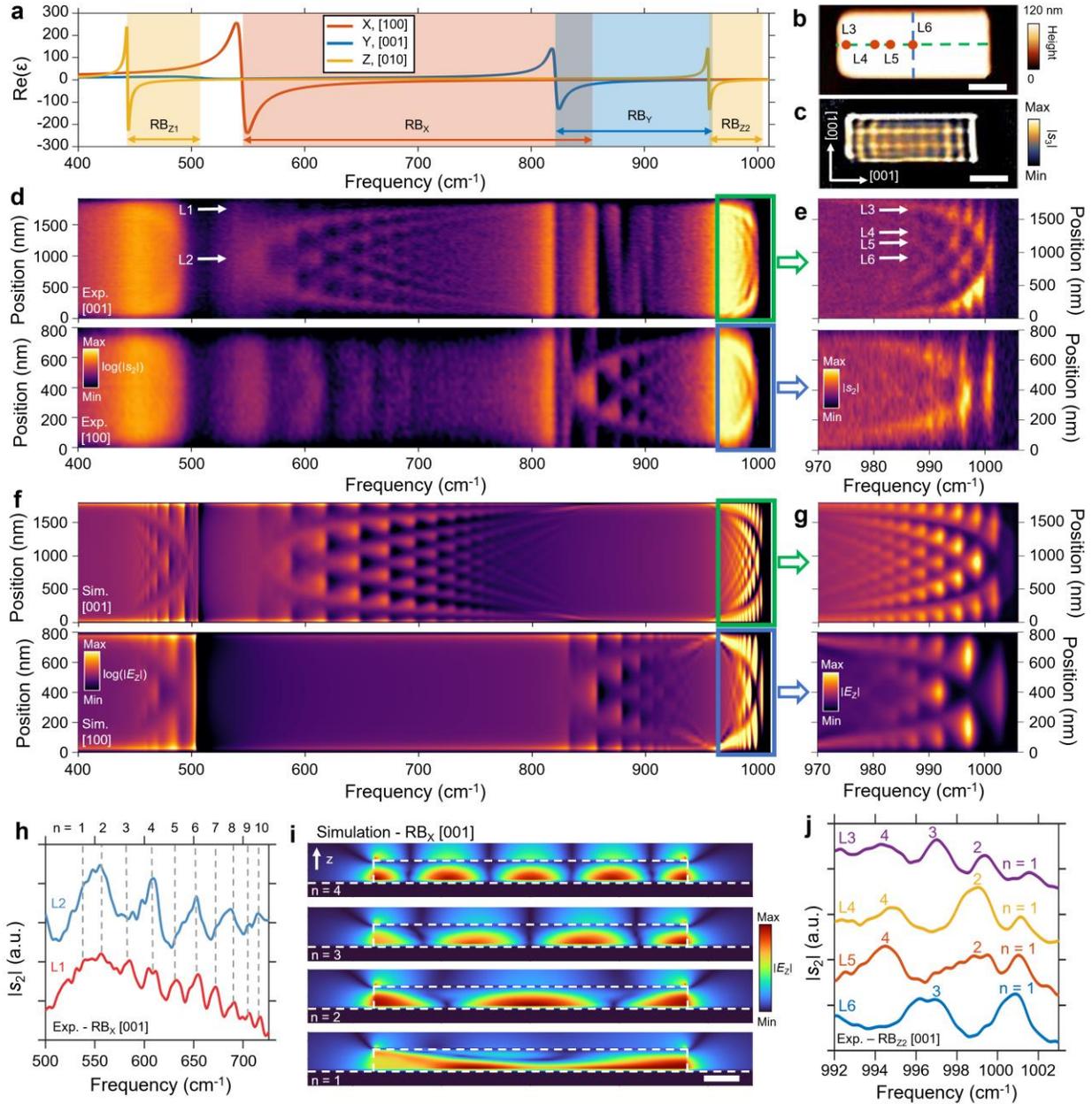

**Fig. 4. Nanospectroscopy of α-MoO₃ nanoribbon on ultrasmooth gold**. **a**, Anisotropic permittivities (real part) of α-MoO$_3$ with the four RBs highlighted. FIR SINS spans over all four RBs and MIR SINS covers the RB$_{Z2}$ band with high spectral resolution. **b**, AFM image of the individual nanoribbon used in SINS analysis. Scale bar: 500 nm. **c**, Simultaneous s-SNOM image

of the nanoribbon in (b) at 980 cm$^{-1}$. The arrows in the AFM image denote the in-plane crystal axes. Scale bar: 500 nm. **d**, FIR SINS line scans along [001] (top) and [100] (bottom) across the nanoribbon. **e**, High-resolution MIR SINS line scans along [001] (top) and [100] (bottom) across the nanoribbon in RB$_{Z2}$. **f**, Simulated spectra along [001] (top) and [100] (bottom). **g**, Zoom-in of the spectra in RB$_{Z2}$ in (**f**). **h**, Spectra taken along [001] in RB$_X$ at the arrow labels L1 and L2 in the top subfigure of (**d**). The FP mode orders are specified with grey dashed lines. **i**, Vertical cross-sections of the simulated electric field |$E_Z$| profiles of the first four FP modes along [001] in RB$_X$. White dashed lines indicate material interfaces. Scale bar: 200 nm. **j**, Spectra taken along [001] in RB$_{Z2}$ at the arrow labels L3 – L6 in the top subfigure of (**e**). The FP mode orders are specified at each peak.

These nanospectroscopy maps can be directly used to quantify the spectral characteristics of different FP modes in the nanoribbon, which are selectively excited and collected as a function of the probe's location on the resonator.[17,30] We present a few spectra in Figures 4h and 4j, which correspond to the arrow-labeled positions in Figures 4d and 4e. For scans taken along the outer nanoribbon edge (Figure 4h), signatures of the $n = 1 – 10$ modes along [001] in RB$_X$ are clearly visible. For scans taken down the center of the nanoribbon, signatures of only the even-order modes ($n = 2, 4, 6…$) are present. Along [001] in RB$_Y$, we observe modes with up to $n = 6$ and similar mode symmetry characteristics at the resonator center and edge (see Supporting Information). The strong dependence of mode excitation on s-SNOM tip position is corroborated with simulated mode profiles, shown in Figure 4i. The odd-numbered modes each have electric field nodes at the resonator center, such that a tip scanning along those positions would not couple with these modes. In contrast, all of the modes have strong electric field components at the resonator edge and can be excited accordingly with s-SNOM.

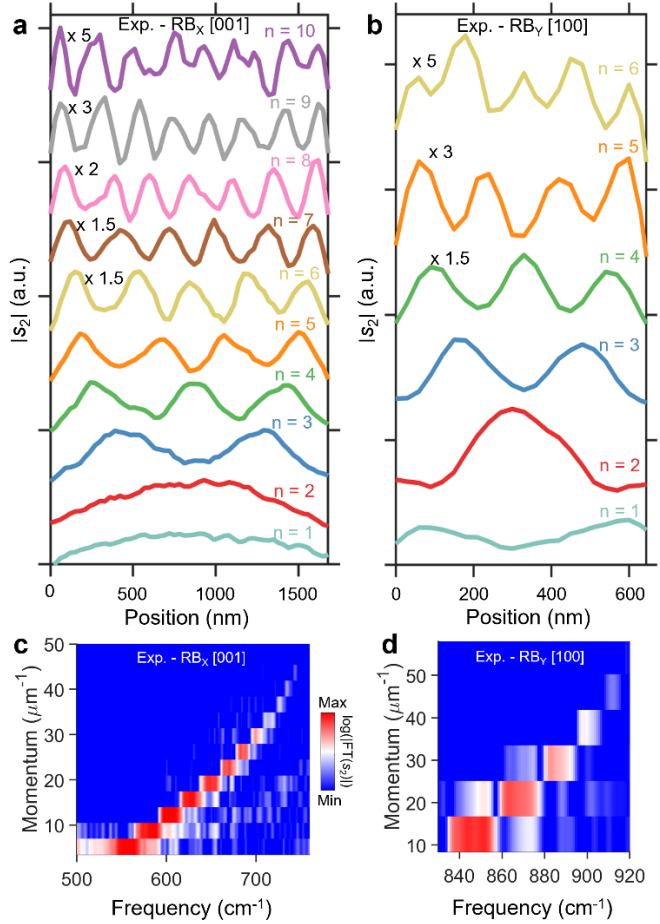

**Fig. 5. Mode profiles of the α-MoO₃ nanoribbon**. **a-b**, Spatial cross-sections of the experimental hyperspectral map along (**a**) [001] in $RB_X$ and (**b**) [100] in $RB_Y$. The data curves are vertically offset for clarity. **c-d**, Fourier transforms of the experimental hyperspectral maps in Figure 4d along (**c**) [001] in $RB_X$ and (**d**) [100] in $RB_Y$. Both plots share the color map in (**c**).

High-resolution MIR spectra, measured at the positions L3 – L6 in Figure 4d, are presented in Figure 4j and display multiple sharp peaks corresponding to the first few FP mode orders in $RB_{z2}$. We quantify the resonator Q-factors by fitting the data with multiple Lorentzian peaks (see Supporting Information) and find that the first, second, and third-order modes have minimum Q-factors of 718 (L3), 548 (L4), and 444 (L6) respectively. These Q-factors are the highest reported in a PhP resonator to date and result from the high crystal quality of the α-MoO₃ nanoribbons, their smooth edges, and the use of an ultrasmooth gold substrate. We hypothesize that the lower Q-factors measured in modes probed at the sample center, compared to those at the sample edge, arise because the modes have relatively strong $E_Z$ components at the sample center that facilitate enhanced radiative out-coupling losses via the probe tip[16]. The Q-factors reported above, which include values taken at the sample center, are therefore conservative values that likely include tip-enhanced radiative losses.

The spatial cross-sections of the FP modes for different mode orders, extracted from the experimental hyperspectral maps, are shown in Figures 5a and 5b. The mode profiles from other RBs are presented in the Supporting Information. FP modes with orders up to $n = 10$ for $RB_X$ and $n = 6$ for $RB_Y$ are clearly observed, even though the broadband light source has much lower power density than that of a typical MIR laser. The fundamental $n = 1$ dipolar mode in Figure 5a is not clearly probed because they are intrinsically weak and poorly couple with the s-SNOM tip (see Figure 4f). A complete experimental hyperbolic PhP dispersion plot can be obtained by applying a Fourier transform (FT) to the high-resolution maps. The FT maps for $RB_X$ and $RB_Y$ bands are shown in Figures 5c and 5d, respectively, and they show FP resonance-induced PhP dispersion quantization that corresponds to FP mode orders $n = 2 - 13$ in $RB_X$ (Figure 5c) and $n = 2 - 6$ in $RB_Y$ (Figure 5d). These trends follow the theoretical non-quantized hyperbolic PhP dispersion curves when considering that the momenta in the quantized dispersion are twice of those in infinite-slab dispersion, due to the standing wave nature of the resonator modes. The $n = 1$ modes are not included as they merge with the DC component. We also estimate the in-plane momenta for different mode orders using the FP condition $2\text{Re}(q)L + 2\varphi = 2\pi n$ (see the Supporting Information) and the results are consistent with the theoretical hyperbolic dispersion.

CONCLUSIONS

In summary, we have demonstrated that FVD-synthesized α-MoO$_3$ micro- and nano-structures are ultrahigh quality phonon polaritonic systems with crystal qualities similar to exfoliated flakes. MIR and FIR s-SNOM studies of α-MoO$_3$ nanoribbons, transferred onto ultrasmooth gold substrates, show that these systems support strong phonon-polariton resonances across four RBs with exceptionally high Q-factors. We anticipate that bottom-up-synthesized oxides will enable new research advances in high-performance and low-loss IR photonic systems along several lines. First, bottom-up synthesis allows for the direct, quick, and scalable production of high-quality polaritonic structures and their study on arbitrary substrates. Second, our FVD uniquely supports control over polaritonic structure morphology and composition, enabling tailoring of chemical doping and composition, defect engineering[46,6], and superstructure assembly[47] through modifications of the materials growth process itself. Third, high-Q resonators based on FVD nanostructures are an attractive and versatile platform for molecular IR sensing and spectroscopy[9,48]. The hyperbolicity of the vdW material can also enable subdiffractional imaging and radiative control applications[10,49,50]. We also anticipate that the mode confinement and low loss featured by our materials make it an ideal platform for studying fundamental PhP light-matter interactions, such as strong and ultrastrong coupling[7].

METHODS

**Sample preparation**

The synthesis of α-MoO$_3$ micro- and nano-structures is based on our previously reported flame vapor deposition (FVD) method[29]. The flame synthesis setup in this work, shown in Figure 1a, has a 6 cm diameter premixed flat-flame burner (McKenna burner, Holthuis & Associates) with CH$_4$ as the fuel and air as the oxidizer. The fuel-to-air equivalence ratio of the flame is fixed as 0.88 with a CH$_4$ flow rate at 1.80 standard liter per minute (SLPM) and an air flow rate at 19.48 SLPM. The equivalence ratio is defined as the ratio of actual fuel/oxidizer molar ratio to the stoichiometric fuel/oxidizer molar ratio. A 4 cm $\times$ 4 cm Mo mesh prepared from Mo wires (0.203 mm in diameter, 99.9 %, Alfa Aesar) is placed over the premixed flame as the solid Mo source. The area density of this Mo source is about 18 mg/cm$^2$. α-MoO$_3$ structures are grown on Si wafer, which is placed above Mo source. The temperature of the Mo source is tuned by adding steel cooling meshes between Mo mesh and burner, and the temperature of the growth substrate is controlled by adjusting the distance between Mo mesh and Si substrate. A K-type thermocouple is used to measure the temperature of Mo mesh and substrate.

α-MoO$_3$ with different morphologies are prepared by tuning the area density of Mo source and the temperatures of Mo source and Si substrate. Higher Mo source temperature (792 $^o$C) is used to facilitate the Mo vapor generation, thus resulting in α-MoO$_3$ microplates (Figure 1d). Lower Mo source temperature leads to smaller nanoribbons (Figure 1e). To demonstrate the growth of α-MoO$_3$ nanoribbons with a higher aspect ratio, the area density of the Mo source is halved (9 mg/cm$^2$), which could reduce the partial pressure of the MoO$_3$ vapor. The substrate temperature is also increased. The driving force for atom attachment is thus lowered and the growth occurs preferentially along the axial [001] direction due to smaller surface energy cost. As a result, the aspect ratio is increased and α-MoO$_3$ nanoribbons are formed (Figure 1f).

All the α-MoO$_3$ samples are prepared by a 5 min deposition. Under such conditions, more than 1 cm$^2$ of the substrate can be uniformly covered by the flame-grown α-MoO$_3$ structures. The as-grown α-MoO$_3$ structures are then transferred onto different substrates (90 nm SiO$_2$/Si and ultrasmooth gold/glass) using low adhesion cleanroom tape (UltraTape 1310) for optical characterizations. The dry transfer method is described as follows. We put a piece of tape on the as-grown sample and gently press on it; we slowly take the tape off and then put it on the target substrate; we gently press and scratch on the tape against the substrate, and then take off the tape slowly. We did not have any sample treatment (such as annealing or solution cleaning) after α-MoO$_3$ transfer. The ultrasmooth Au substrate (Au thickness: 100 nm) is template stripped from a Si wafer (Platypus Technologies). The coverage of as-transferred micro- and nano-structures on the target substrate is also at the centimeter scale.

The exfoliated α-MoO$_3$ flakes for phonon lifetime comparison are tape exfoliated and transferred on 90 nm SiO$_2$/Si substrates from commercially available bulk single crystals (2D Semiconductors).

**Near-field optical imaging and nanospectroscopy**

PhPs in the α-MoO$_3$ nanostructures are imaged with a commercial s-SNOM system (Neaspec) and a tunable CO$_2$ laser (Access Laser). Sharp metal-coated tips, tapping at ~280 kHz, are used to probe near-field signals in the pseudoheterodyne interferometric mode. The collected light is detected with a liquid-nitrogen cooled mercury-cadmium-telluride (MCT) detector (Kolmar KLD-0.1). The s-SNOM imaging is done at ambient conditions. The third-order harmonic near-field signal $s_3$ is presented and analyzed.

Synchrotron infrared nanospectroscopy (SINS) measurements were taken at Beamlines 2.4 and 5.4 at the Advanced Light Source (ALS), Lawrence Berkeley National Laboratory[42,43]. The FIR SINS system is a commercial s-SNOM system (Neaspec) interfaced with the synchrotron light source and a liquid-helium cooled Ge:Cu detector, covering a range of 320 – 1500 cm$^{-1}$. The MIR SINS system integrates the synchrotron light source with a customized, commercial AFM (Bruker Innova), liquid-nitrogen cooled MCT detector, and a commercial Fourier-transform infrared spectrometer (Thermo-Scientific, Nicolet 6700), covering a range of 700 – 5000 cm$^{-1}$. The SINS measurements are done at ambient conditions. The second-order harmonic near-field signal $s_2$ (amplitude denoted by $|s_2|$ and phase $\arg(s_2)$) is used for hyperspectral mapping, with the spectral resolution set at 4 cm$^{-1}$ for the FIR and 0.5 cm$^{-1}$ for the MIR measurements. The spatial step is 30 nm for the FIR linescans, 37 nm for the MIR [001] linescan, and 41 nm for the MIR [100] linescan. The hyperspectral maps in Figure 4 are normalized to the reference nanospectrum of bare ultrasmooth gold[43] and the maps are plotted with interpolation. A Happ-Genzel apodization function is applied in our Fourier transform computation.

AUTHOR INFORMATION

**Corresponding Author**

*Jonathan A. Fan

jonfan@stanford.edu

Spilker Building, Room 307

Stanford University

348 Via Pueblo

Stanford, CA 94305-4088 USA

(650) 723-0278

**Author contributions**

J.F., X.Z. and T.H. supervised the project. S.Y. and J.F. conceived the project idea. Y.J. prepared FVD samples. Y.J. and X.S. performed basic material characterization. S.Y. performed s-SNOM imaging. S.Y., H.B. and S.C. performed synchrotron nanospectroscopy. S.Y. performed experimental data analysis. S.Y. and J.R. performed simulation and theoretical analysis. M.H. supported s-SNOM imaging instrumentation. H.Y. prepared exfoliated flake samples. All authors analyzed the results, wrote, and edited the manuscript. S.Y. and Y.J. contributed equally to this work.

NOTES

The authors declare no competing financial interest.

ACKNOWLEDGMENT

The authors acknowledge support from the National Science Foundation (NSF) under award no. 1804224, the Air Force Office of Scientific Research (AFOSR) Multidisciplinary University Research Initiative (MURI) under award no. FA9550-16-1-0031, AFOSR under award no. FA9550-18-1-0070, and the Packard Fellowship Foundation. The s-SNOM measurements were supported by the Department of Energy (DOE) "Photonics at Thermodynamic Limits" Energy Frontier Research Center under grant no. DE-SC0019140. X.Z. acknowledges support from the NSF EFRI-DCheM program under award no. SUB0000425. J.R. acknowledges support from the Department of Defense through the National Defense Science and Engineering Graduate Fellowship Program. This research used resources of the Advanced Light Source, a U.S. DOE Office of Science User Facility under contract no. DE-AC02-05CH11231.